\documentstyle[aabib,epsf,psfig]{l-aa}
\newcommand{\Kms}{\mbox{km~s$^{-1}$}}
\newcommand{\msun}{\mbox{${\rm M}_\odot$}}

\newcommand{\rsun}{\mbox{${\rm R}_\odot$}}
\newcommand{\Zytkow}{$\dot{\mbox{Z}}$ytkow }

\newcommand{\vsun}{\mbox{${\rm v}_\odot$}}
\newcommand{\st}{\mbox{$\star$}}
\newcommand{\AU}{\mbox{{\rm AU}}}

\begin{document}
\thesaurus{03.13.4;
           05.03.1;
           08.02.1;
           08.05.3;
           08.02.7;
           10.07.2}
\title{Star Cluster Ecology II:}
\subtitle{Binary evolution with single-star encounters}
 \author{Simon F.\ Portegies Zwart$^{1, 2}$,
         Piet Hut$^3$,
         Stephen L.\ W.\ McMillan$^4$ \&
         Frank Verbunt$^2$}
\institute{$^1$ Astronomical Institute {\em Anton Pannekoek}, 
                Kruislaan 403, NL-1098 SJ Amsterdam \\
           $^2$ Astronomical Institute, Utrecht, Postbus 80000, 
                NL-3508 TA Utrecht \\
           $^3$ Institute for Advanced Study,
                Princeton NJ, USA \\
           $^4$ Department of Atmospheric Science, Drexel University,
                Philadelphia, PA 19104}

\date{}
\maketitle
\markboth{Simon F.\ Portegies Zwart {\em et al.}}
         {Star Cluster Ecology II}

\begin{abstract}
 
Three-body effects greatly complicate stellar evolution.  We model the
effects of encounters of binaries with single stars, based on
parameters chosen from conditions prevalent in the cores of globular
clusters.  For our three-body encounters, we start with a population
of primordial binaries, and choose incoming stars from the evolving
single star populations of paper I.  In addition, we study the
formation of new binaries through tidal capture among the single
stars.  In subsequent papers in this series, we will combine stellar
evolution with the dynamics of a full $N$-body system.

\end{abstract}

\section{Introduction}
\label{sec_71}
In a dense stellar system, such as an open or a globular cluster or a
galactic nucleus, encounters between individual stars and binaries can
affect the dynamical evolution of the system as a whole.  As a first
step toward a more detailed description of such a stellar system,
Portegies Zwart et al. (1997, hereafter paper~I)\nocite{pzhv97}
modeled encounters and encounter products in a population of evolving
single stars.  In these calculations the stellar number density was
held constant in space and time, as a first approximation to the
collapsed core of a globular cluster.

As a second step we present in this paper the evolution of binaries in
such a core, as follows.  We add a single binary to the stellar system
and follow its evolution until the binary is destroyed or ejected, or
until the simulation terminates.  This procedure is repeated to build
up a large number of binary evolution histories.  Thus we ignore in
this paper encounters between binaries; we also ignore encounters
between binaries and the products of binary evolution.  A similar
experiment was performed by \cite*{dav95}, but he neglected the
detailed internal evolution of close binaries (see also the work of
\cite{egg96}).

We consider both primordial binaries, formed simultaneously with the
rest of the cluster, and tidal binaries formed later from encounters
between single stars.  Each binary evolves both through internal
processes (e.g.\ the evolution of its components, stellar wind, mass
transfer, etc.) and due to encounters with single stars.

The main purpose of this study is to gain a better understanding of
the interactive processes between dynamics, stellar evolution and
binary evolution.  Our goal for the next episode in this series is to
replace the integration of only 3-body systems with the full N-body
evolution of the stellar system as a whole.  The first pioneering
experiments in this respect are performed by \cite*{aar96}.

The choices for the binary population in the cluster core, and the
treatment of the internal evolution of the binaries and of encounters
between stars and binaries are described in sect.~\ref{sec_72}.  The
results of our calculations are presented in sect.~\ref{sec_73} and
\ref{sec_74}, for high- and low-density clusters, respectively.  A
brief preview of future work is given in sect.~\ref{sec_75}.

\section{Methods}
\label{sec_72}

In this section we describe the various steps that take place in a
single model calculation in which a population of binaries is evolved
in a background of evolving single stars.  In the present paper, we
consider the limit of a low binary fraction: we model encounters
between single stars and binaries, as well as those between single
stars and single stars, but we neglect the occurrence of binary-binary
encounters.  The reasons for this choice are twofold.  First, it is a
good approximation to the core of a globular cluster in the asymptotic
post-collapse regime, in which most primordial binaries have already
been ``burned up'' (\cite{mh94}).  Second, it brings about a
considerable simplification, in that it allows us to decouple the
evolution of the single stars and the binaries.

As a first pass through the history of the star cluster, we follow
only the single stars.  We model the collisions between all types of
single stars: both the primordial single stars, as well as the newly
formed collision products.  We thus model their internal stellar
evolution, as well as their non-linear dynamical collision history
(non-linear in the sense that doubling the density of stars will
more than double the number of collisions per star, given the higher
chance for collisions with and between the heavier collision products
in the form of blue and yellow stragglers).

As a second pass, we model the collisions between single stars and
binaries.  As mentioned above, we limit ourselves to a strictly linear
treatment, in the sense that each binary can encounter any of the
single stars (primordial ones as well as stragglers), but is not
allowed to meet another binary.  Therefore, there is no need to evolve
the binaries simultaneously.  Instead, we simply decide how many
binaries we would like to follow in total, and then evolve one binary
at a time, until we have reached our quota.

Given the separability of the evolution of single stars and binaries,
we can simply take the results of paper I to describe the former.  For
the binaries, we must (1) follow the internal changes caused by the
stellar evolution of the individual binary member stars, (2) keep
track of changes in the binary orbit caused by mass loss from, and
mass exchange between, the two stars, and (3) model encounters between
binaries and single stars.

Our overall approach is as follows.  We start with a single-star
environment taken from paper I, as described in sect.~\ref{sec_ssenv}.
In our ``low-density'' environment (model S below), collisions are
rare and are modeled from the outset.  However, in our
``high-density'' environment (model C), collisions are allowed only
after a specific time $t_{\rm cc}$ (cc for core collapse), effectively
using a step function to model the fact that collisions become
important only during the later stages of dynamical evolution, around
the time of core collapse.

We start the calculations in this paper by selecting one binary, as
outlined in sect.~\ref{sec_selecb}.  There are two possibilities,
depending on whether this binary is primordial, formed together with
the rest of the cluster, or a product of tidal capture, and thus
formed at a later time.  A primordial binary is evolved in isolation
until the time $t_{\rm cc}$ (sect.~\ref{sec_evb}), at which point it
starts to interact with the single stars.  In contrast, a tidal binary
can only be formed at a later time, in our model, and we thus have to
update our single-star environment to reflect this fact.  In either
case, we are then ready to implement dynamical three-body encounters
between binaries and single stars chosen randomly from the appropriate
distribution function.

In sect.~\ref{sec_selb}, we describe the selection of these
three-body scattering events.  In a nutshell, we choose a time step,
and determine whether or not the binary encounters a single star in
this time step.  If an encounter is indicated, we carry out the
scattering operation, and we update both the binary and the stellar
system, as detailed in sect.~\ref{sec_encb}.  We then choose a new
time step, and repeat our Monte-Carlo selection procedure.  This
process is continued until the binary is destroyed or until the end of
the dynamical calculation, whichever occurs first.  We then select a
new binary, repeat the whole binary evolution procedure, and continue
until we have followed the required number of binaries.  A list of
terms used in our scattering experiments is provided in
sect.~\ref{sec_term}.

\subsection{The single star environment}
\label{sec_ssenv}

We consider two stellar environments, one with a Salpeter type mass
function and a moderate density which we call model~$S$, and one with
a mass function that is strongly affected by mass segregation and with
a high density, appropriate for a post collapsed core, which we call
model $C$.  The relevant parameters of these models are given in
Tables~\ref{Tab_initb} and \ref{Tab_init} (see also paper~I).

We evolve the stellar system with single stars and collisions between
them as described in paper~I.  Two stars are assumed to merge into a
single object if they approach each other within a distance equal to
the sum of their stellar radii (note that for paper~I, twice this
value was used; we discuss our reasons for making a different choice
here in sect.~3.3).  Binaries that pass each other at a slightly
larger distance experience a tidal encounter, which might directly
lead to coalescence. If the tidal encounter results in the formation
of a binary, the orbital parameters of this binary are stored together
with the mass and evolutionary state of the two stars in a data base
of tidal binaries.  Binaries of this type are evolved in a later pass,
as described in sect.~\ref{tidals}.

The stellar content of the cluster is stored in binned form (see
sect. 3.4 of paper~I) at regular time intervals $\Delta t$.  The
appearance of the stellar system between two stored moments $t$ and
$t+\Delta t$ is determined by interpolating the number densities of
the various stellar masses, radii and types linearly in time between
those two instances.  The environment of single stars is not affected
by the binaries that are evolved within it; a collision product formed
in a dynamical encounter between a single star and a binary is not
added to the environment, and a single star that is destroyed (or
exchanged) by an encounter is not discarded from the environment.
With each bin, a single number is associated to indicate the number of
stars at a given time interval, within the ranges in mass and radius
corresponding to that bin.  The identity of the individual stars is
thus lost, in order to reduce the storage requirements for the
single-star data base.

Whenever a binary is taken as a candidate for interaction with single
stars, at time $t$, the single star environment is retrieved from the
data base, and interpolated between the two storage times that
straddle the time $t$.  Having thus synchronized the single stars with
the binary, we can determine whether an three-body encounter occurs
(see sect.~\ref{sec_selb}).  If an encounter occurs between one of
the retrieved stars and the binary, the current evolutionary state of
the star has to be known.  In the recovery process from the stored
stellar mass, radius and type of a star, the individual age of the
star is not known.  In most cases, when the star is not affected by
earlier collisions this poses no problem: the stellar age is equal to
the age of the stellar system.  Things are more complicated in case
the star has been formed as a collision product.  This implies that
the star is relatively young compared to the stellar system.  We
assign an age for this rejuvenated star with the following procedure.

 From the mass, radius and stellar type of a retrieved star its age can
be uniquely determined.  When a star is more massive than the other
cluster members in the same evolutionary state, its implied stellar
age is less than the age of the system as a whole.  In order to know
exactly how much smaller, we would have to know the identities of the
progenitor stars, and the amount of mixing that has occurred during
the merging process.  The latter is still a matter of considerable
debate, and the former is something we do not have access, since our
binning erases specific information pertaining to individual stars.
Therefore, we have adopted a simple recipe: we assume that the amount
of mass in excess to a normal star in the same evolutionary state is
accreted onto the rejuvenated star (see paper~I for a discussion of
this treatment).

This recipe will make some errors.  For example, let us consider a
globular cluster with a turn-off mass of $0.8\msun$.  In this
environment, if two $0.5\msun$ stars collide, they form a blue
straggler with a total mass of $1\msun$.  Our recipe will treat this
system in the same way as a similar process in which a $0.2\msun$ star
would have collided with a $0.8\msun$ star.  In the latter case, the
addition of $0.2\msun$ will only briefly delay the turn-off star in
its ascent onto the giant branch.  In contrast, the actual system, in
which two equal-mass stars have collided, can be expected to live
considerably longer, since both stars were still far from exhausting
the hydrogen in their cores.  Fortunately, in most cases the error
made by our approximate treatment is relatively small, since the
high-mass merger remnants typically have much smaller lifetimes than
the age of the system (see Fig.~6 in paper~I).

\subsection{Initialization of the binaries}
\label{sec_selecb}

We consider the evolution of both primordial binaries and tidal
binaries.  Here we treat the initialization of each type of binary in
turn.

\subsubsection{Tidal binaries}
\label{tidals}

To determine the formation properties of tidal binaries, we must first
determine, at any given time, whether any two stars in the stellar
system are about to undergo an encounter that is sufficiently close to
offer the possibility of collisions or tidal capture.  The calculation
required is similar in many respects to the calculations carried out
in paper~I, where we studied encounters between single stars leading
to collisional merging.  In that paper, stars were assumed to merge
when the distance of closest encounter $d$ between their centers was
less than twice the sum of their radii: $d \le 2(R_1 + R_2)$, where
$R_i$ is the radius of star $i$.  For purposes of computing tidal
capture rates, we must consider larger $d$ values.  As an upper limit
for $d$, we have chosen $d_{\rm max} = 5(R_1 + R_2)$.  This is a safe
upper limit for conditions under which tidal capture may occur.

As described in paper~I, the velocity distribution of stars with mass
$m$ is given by a Maxwellian, and therefore the distribution of
relative velocities of two stars chosen at random are given by
another Maxwellian, which we denote $F(v)$.  The stars that are
involved in an encounter are not chosen randomly, however: stars with
lower relative speeds have considerably higher probability of becoming
involved in an encounter.  We take this into account by choosing the
relative stellar velocity from a distribution $P(v)$ weighted by the
encounter rate $\sigma v$, where $\sigma$ is the cross section for an
encounter to occur:
\begin{equation}
        P(v) dv = \sigma v F(v)dv.
\label{eq:rel_vel}\end{equation}
The velocity distribution $P(v)$ (Fig.~1) depends on $d_{\rm max}$,
through the dependence of the cross section $\sigma_{\rm
max}=\sigma(d_{\rm max})$ on $d_{\rm max}$.

After choosing the relative encounter velocity from Eq.~\ref{eq:rel_vel},
the cross section $\sigma_{\rm max}$ for any encounter leading to
stellar collision or tidal capture is
\begin{equation}
        \sigma_{\rm max} = \pi d_{\rm max}^2 \left( 1 + 2G \frac{M_1 + M_2}
                               {v^2 d_{\rm max}}
                               \right),
\label{eq:sigma}\end{equation}
where $M_i$ is the mass of star $i$ and $v$ is the
relative velocity between the two stars at infinity.

The probability distribution for an encounter is linear in $\sigma$,
by definition.  Therefore, we can choose a random value for $\sigma$
 from a flat distribution between 0 and $\sigma_{\rm max}$.  We can then
obtain the distance of closest approach $d$ by inverting $\sigma(d)$
(see Eq.~\ref{eq:sigma} with $d$ substituted for $d_{\rm max}$, and
$\sigma$ substituted for $\sigma_{\rm max}$) to obtain $d(\sigma)$.

\begin{figure}
\hspace*{0.5cm}
\psfig{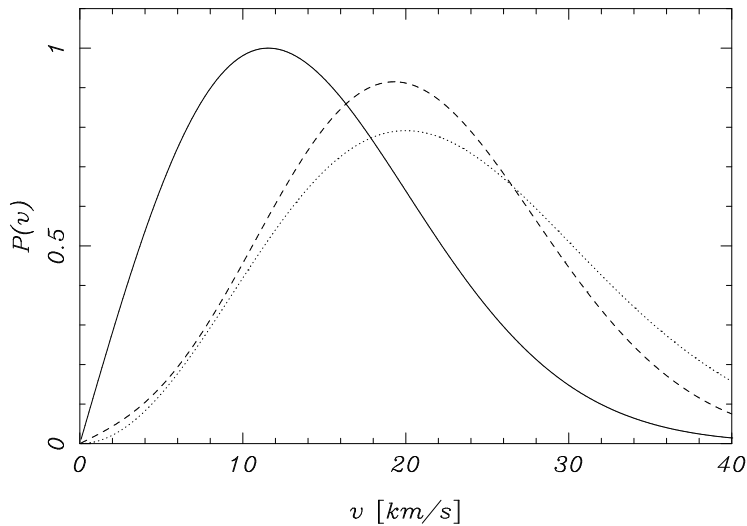}
\caption[]{Velocity distributions $P(v)$ for ``interesting''
encounters (i.e. leading to a physical stellar collision or to a
significant binary perturbation) to occur, for the cases (a) two
identical 1~\msun, 1~\rsun\ stars (solid line), and (b) a binary with
a total mass of 1~\msun\ and a semi-major axis of 1~\AU\ encountering
a star with a mass of 1~\msun\ and a radius of 1~\rsun (dashed line).
For comparison, the dotted line indicates the Maxwellian velocity
distribution with a dispersion of $\sqrt{2} \vsun$
(Eq.~\ref{Eq_vdisp}).  The solid curve is arbitrarily normalized to a
maximum value of unity; the areas under all three curves are the same.
}\label{fig_relvel}\end{figure}

The amount of energy deposited into oscillatory modes in the two stars
at the moment of periastron passage is computed using the method
described by Press \& Teukolsky (1977, and refined by \cite{lo86},
\cite{mmt87}, \cite{rka87}; see also Mardling
1995a,b)\nocite{mar95a}\nocite{mar95b}\nocite{pt77}, where each star
is taken to act upon the other as a point mass.  The amount of energy
dissipated in the tidal interaction is computed using the
semi-analytic formulae provided by \cite*{pzm93}.  These equations
give, as a function of primary and secondary mass, radius and
polytropic index $\gamma$ the total energy $E_{\rm tide}$ dissipated
in the first periastron passage.  For the computation of the amount of
energy dissipated we use a polytropic index of $\gamma=3/2$ for all
stars.

The final energy of the two-body system is calculated by subtracting
the dissipated energy from the kinetic energy at infinity before the
encounter.  If the total energy is positive, no binary is formed, and
the encounter is discarded.  If the total energy is negative, the
semi-major axis and eccentricity of the newly formed binary are
calculated from the total energy and distance of closest approach $d$,
assumed equal to the periastron distance of the new orbit.  We then
model the subsequent (partial) tidal circularization of the binary as
an instantaneous process, as follows.  We let the binary reduce its
eccentricity under conservation of angular momentum until either of
two types of outcome takes place.  In some cases, the binary will
become circular before the periastron distance has reached a value of
four times the radius of the largest component.  In other cases, this
criterion will be reached for non-zero eccentricity.  In both cases we
freeze the binary, neglecting the remaining tidal interaction in the
latter case (until stellar evolution effects increase the radii of the
stars).

If one of the stars fills its Roche lobe after the binary has been
circularized, the stars are merged and the encounter is classified as
a collision.  After the formation of a tidal binary, both stars are
removed from the system of single stars and the orbital and stellar
parameters of the binary, as well as the time of capture, are stored.
Later, in the second pass in which we consider tidal binaries as
candidates for three-body encounters with single stars, this stored
information is retrieved to evolve the tidal binaries in the
collisional environment.

\subsubsection{Primordial binaries}
\label{sec_prim}
The primordial binaries are initialized at zero age with distribution
functions for the mass of the primary, the mass ratio, the semi-major
axis and the eccentricity.  All primordial binaries contain two
main-sequence stars.  The mass of the primary $M$ is taken from the
mass function for the single stars (see paper~I, sect.\ 3.1.1).  After
the primary mass is chosen, the mass $m$ of the secondary is selected,
between the minimum mass of 0.1 \msun\ and the mass of the primary,
 from a specified probability distribution $\Phi(q)$, where the mass
ratio $q=m/M$ is defined to be smaller than unity (see Table~\ref{Tab_initb}).

\begin{table}
\caption[]{Initial conditions for the primordial binaries for the low
density model S, with a Salpeter primary mass function.  The
parameters are listed in the first column, the distribution function
in the second, and the lower and upper limits are given for the models
$C$ and $S$.  A primordial binary that is not detached at zero age is
rejected.  Note that in model C the initial parameters listed here are
substantially modified to take into account the effects of dynamical
evolution prior to the start of the simulation at time $t_{cc}$.  As a
result, the initial mass function actually used in that model is
considerably flatter than Salpeter.} 
\begin{center}
\begin{tabular}{l|llr} \hline
       & & \multicolumn{2}{c|}{Models $S$ \& C} \\
Symbol       & Function & min & max  \\ \hline
$M$         &$\Psi(M) \propto M^{-2.35}$& 0.1\msun  & 100\msun \\
$q$         &$\Phi(q) =$ {\em const.}       &$0.1\msun/M$& 1  \\
$e$         &$\Xi(e)  = 2e$               & 0  & 1     \\
$a$         &$\Gamma(a) \propto 1/a$ & 1\rsun  &10\AU  \\ \hline
\multicolumn{2}{l}{criterion for survival:}
            &\multicolumn{2}{l}{detached}\\
\end{tabular}
\end{center}
\label{Tab_initb}\end{table}

This procedure is biased, in that only the primary has been selected
 from the single star mass function, without any regard to the mass of
the secondary. In model $C$, with a strongly segregated mass function,
this treatment is not very realistic.  Due to the large selection effects
in observational analysis, it is not clear how we can best correct for
this bias.  Therefore we have adopted the following simple algorithm.
We require that distribution of total masses of the resulting binaries
follow the same probability distribution as the masses of the single
stars. For the distribution of the single stars we use the mass
function for the single stars in model $C$ (see paper~I). A rejection
technique is used to enforce this distribution, after first selecting
a large number of candidate binaries.

The dispersion velocity of the center of mass of the binary is chosen
according to equipartition:
\begin{equation}
        v(M+m) = \sqrt{1\msun\over M+m} \vsun,
\end{equation}
where \vsun\ is the three-dimensional velocity dispersion for a
1\msun\ star, for which we choose (see also paper~I):
\begin{equation}
\vsun = \sqrt{3} \times 10 \Kms \approx 17.3 \Kms. 
\label{Eq_vdisp}\end{equation}
The initial eccentricity distribution for a binary system is chosen to
be thermal; $\Xi(e) = 2 e$ (see e.g.\ Duquennoy \& Mayor 1991),
\nocite
{dm91}
with $0 \leq e < 1$.
The initial semi-major axis is chosen from a distribution which is flat
in $\log a$ (Kraicheva et al.\ 1978) \nocite{kpt+78}
with $a_{\rm min} \leq a \leq a_{\rm max}$.
We choose $a_{\rm min} = 1~\rsun$ and $a_{\rm max} =$~10~\AU.
After choosing the binary parameters we test whether
the binary is detached; if not, one of the stars is filling its
Roche-lobe, the binary is rejected and new parameters are chosen.

\subsection{Internal evolution of the binaries}
\label{sec_evb}
Our treatment for the internal evolution of a binary, including mass
loss via stellar wind, mass transfer from one component to another,
and other physical processes, follows the detailed description of
model $AK$ from \cite*{pzv96}.

The description in \cite*{pzv96} was mainly devoted to high-mass
binaries and the treatment for accreting white dwarfs was rather
imprecise.  In the present context, accreting white dwarfs are quite
common and a more refined treatment is appropriate.  In observed novae
the amount of mass accreted onto the white dwarf is generally smaller
than the amount of mass ejected in a fast stellar wind (see
e.g.~\cite{lp92}).  Mass dumped onto the white dwarf is not directly
accreted by it, but rather temporarily stored a circumstellar disc or
envelope.  For a disc formed around a white dwarf after coalescence
with a less massive star, we assume that the dwarf accretes at a rate
equal to 1\%\ of the Eddington limit and loses mass at a rate equal to
the Eddington limit until the disc is lost.  In the case of mass
transfer onto a white dwarf from a binary companion, the disc/envelope
accretes at a the Eddington limit and the white dwarf accretes at 1\%\
of Eddington.  When the accumulated mass in the envelope exceeds 1\% of
the total mass of the white dwarf a thermonuclear explosion expels the
envelope.  If the distance at pericenter is smaller than 1~\rsun, we
consider the expelled mass to be discarded in a common envelope,
decreasing the binary orbital separation.  Otherwise, the material
lost by the envelope is considered to be blown away in an isotropic
stellar wind.

We define the internal evolution time limit of the binary $\tau_{\rm
evb}$, as the time difference between the current time and the time at
which one of the two stars will evolve into another evolutionary
state.  The categories we have chosen as signifying different
evolutionary states are: main sequence, Hertzsprung gap, (sub)giant,
horizontal branch, supergiant, as well as Wolf-Rayet and helium
stars; see \cite*{pzv96}.

\subsection{Initialization of three-body scattering events}
\label{sec_selb}

The encounter rate between a given binary and all single stars in the
stellar system is computed as follows.

We bin the stars in intervals of mass and radius (see sect.\ 3.4 in
paper~I).  Within a bin, all stars share the same mass, radius and
velocity dispersion.  For each bin, we compute the probability for an
encounter between stars from this bin and the binary.

The encounter probability between a representative star from a bin and
the binary is determined through a procedure, similar to the one
described above, in sect.~\ref{tidals}.  There, we used for the
maximum distance of closest approach, $d_{\rm max}$, a value $d_{\rm
max} = 5(R_1 + R_2)$.  Here, we can take the same expression, but we
have to modify it in three ways.

First, we take as the ``radius'' of the binary its semi-major axis.
Second, we can expect the need for a somewhat larger scaling factor,
since we are interested in relatively mild interactions as well, in
which the eccentricity may still be perturbed sufficiently to affect
the binary noticeably.  Finally, we must take into account the
possibility that the incoming star is much heavier than the sum of the
binary member masses, in which case a more distant encounter can still
cause a significant perturbation on the binary.  From the distance
dependence of tidal interactions, it follows that the latter effect is
proportional to the cube root of the mass of the single star.

After extensive trial runs, we found that the following expression of
maximum value for the closest encounter distance was a safe choice:
\begin{equation}
d_{\rm max} = 7 (a + R_3) 
              \max\left(1, \left({M_3 \over M_1+M_2}\right)\right)^{1/3}.
\end{equation}
Here $R_i$ and $M_i$ are the radius and mass of the binary members (1
and 2) and of the incoming star (3), and $a$ is the semi-major axis of
the binary orbit.

Following the procedure of sect.~\ref{tidals}, with the above
expression for $d_{\rm max}$, we can determine the partial cross
section $\sigma_{sb}(v)$ for encounters between the binary and any of
the stars from a given bin that lead to collisions or a non-negligible
binary perturbation.  Note that this partial cross section is still a
function of the velocity $v$.  In sect.~\ref{tidals} we choose the
velocity through a rejection technique.  In the present case, we have
to integrate the equivalent of Eq.~\ref{eq:rel_vel}, to obtain the
encounter rate $\Gamma_{sb}$ for any star from bin $s$ with the
binary, given by
\begin{equation}
\Gamma_{sb} = n_s \langle \sigma_{sb}v_{sb} \rangle,
\end{equation}
where the brackets indicate averaging over the velocity distribution
which is taken to be a Maxwellian, and $n_s$ is the number density of
stars from bin $s$.

The total encounter rate $\Gamma$ is obtained by summing over all
single-star bins ($N_{bin}$ in total):
\begin{equation}
        \Gamma = \sum_{s=1}^{N_{bin}} \Gamma_{sb} 
                        \equiv {1\over \tau_{\rm enc}}.
\label{renc_b}\end{equation}
Here $\tau_{\rm enc}$ is introduced as a characteristic time interval
between successive interesting encounters, between the given
binary and all single stars.

At the beginning of each time step, we determine the distribution of
stars over the bins and the number densities of stars in each bin,
interpolating from the previous computations with single stars and
compute the times scales for stellar evolution $\tau_{\rm ev}$ and
collisions $\tau_{\rm enc}$.  Here $\tau_{\rm ev}$ is the time
difference between the current time and the next time at which the
internal evolution of the binary is scheduled to be updated.  The task
of computing the sum over all bins $i$ in Eq.~\ref{renc_b} is less
daunting as may appear at first sight, as many bins contain no
stars. This is illustrated in Fig.~\ref{fig_rencfC} below.  The time
step is then calculated as
\begin{equation}
\delta t = \min (0.05\tau_{\rm enc}, \tau_{\rm ev}, \tau_{\rm evb}),
\end{equation}
to ensure that changes in the stellar population and in the binary are
followed with sufficient resolution.

A rejection technique is used to keep track of collisions, as follows.
We choose a random number between 0 and 1. If this number is larger
than $\Gamma \delta t$, we conclude that no collision has occurred. We
evolve all stars and the binary over a time interval $\delta t$, and
continue with the next step.  If the random number is smaller than
$\Gamma \delta t$, a collision has occurred.  In calculating the sum
(Eq.~\ref{renc_b}) over the bins, we keep track of the partial sum
after addition of each bin $i$, where the partial sum ranges over bins
$1, 2, ..., i$.  The first bin for which this growing partial sum
exceeds the random number identifies the bin involved in the
collision.

More than one stellar type may be found in a single mass-radius bin.
For example, white dwarfs with a mass close to the Chandrasekhar limit
and neutron stars can share a bin, and so can Thorne-\Zytkow\ objects
and massive super giants.  As mentioned earlier, the identities of
individual stars are lost within each bin, but we keep track of the
number of stars of each type.  The actual type of star that is
involved in the encounter is selected randomly from the bin with a
probability proportional to the number of stars of each type in the
bin.

\subsection{Execution of three-body scattering events}
\label{sec_encb}

Once a single star is selected, the initial conditions for the
encounter with the binary are determined in a way similar to the
selection procedure described in sect.~\ref{tidals}.  In particular,
the relative encounter velocity between the binary and the single star
is chosen randomly from the following distribution, corresponding to
Eq.~\ref{rel_vel} (see Fig.~\ref{fig_relvel}):
\begin{equation}
        P_{sb}(v) dv = \sigma_{sb} v F(v)dv.
\label{rel_vel}\end{equation}
The maximum impact parameter $p_{\rm max}$ follows from
the defining relation $\sigma_{sb} = \pi p_{max}^2$, and the actual
impact parameter value $p$ is determined randomly from a distribution
flat in $p^2$.

After the relative velocity at infinity and the impact parameter have
been chosen, the remaining parameters are taken randomly from their
appropriate distributions (see \cite{hb83} and \cite{mh96} for
details).  The three-body scattering event is then simulated, through
explicit orbit integration, using the scatter3 module of the Starlab
package developed by \cite*{mh96}, which automatically determines the
type of outcome of the scattering experiment.

If the incoming star emerges from the scattering process unchanged,
leaving the binary behind in a new stable orbit, the scattering event
is classified as a preservation encounter.  If the outgoing star was
originally a binary member, we designate the event an exchange
reaction.  In the case that all three stars emerge unbound, we speak
of an ionization event; this can only occur when the total energy of
the three-body system is positive.  For negative total energy (that
is, when the binary binding energy is larger than the kinetic energy
at infinity of the relative orbit), a resonance scattering may occur,
in which the three stars form a temporary bound state; such an
interaction is characterized by multiple close encounters.

In addition to these outcomes, which can all occur in the point-mass
limit, there are additional channels in our case, when stars have
finite sizes.  One possibility is that all three stars collide,
leaving a single remnant.  If two of the three stars collide, the
remaining star can be either be unbound and escape, in which case we
speak of a collision-ionization event, or it can remain bound in a new
binary system with the merger remnant, in which case we speak of a
collision-binary-formation.

We use the following notation to classify the outcome of a scattering
event.  The numbers 1, 2, and 3 indicate the primary and secondary of
the binary, and the incoming star, respectively.  Closed brackets are
used to show that two stars form a binary.  Braces designate a merger
remnant, with the numbers within brackets indicating which of the
stars were involved in the merger process.  For example: an encounter
that results in the formation of a new binary with the unaffected
field star (3) in a binary with the merger product of the primary (1)
and the secondary (2) is written as: $(\{1, 2\}, 3)$; a triple merger
is written as $\{1, 2, 3\}$; and ionization of the binary, leaving the
third star unaffected, is written as $1, 2, 3$.

All encounters are computed in the center-of-mass frame of the
three-body system.  If, after the encounter, the center-of-mass
velocity of the binary (assuming that a binary remains after the
encounter) exceeds the escape velocity of the stellar system, then the
binary is considered to escape from the system.  Following the usual
approximation, we take the escape velocity to be twice the velocity
dispersion.

A soft binary, with a binding energy much smaller than the typical
kinetic energy of a single star, is likely to become ionized through
repeated encounters.  This process of ionization often occurs after a
large number of weak encounters, in which the binary is gradually
pushed to higher and higher ``energy levels'' (wider orbits).  To
model this process in detail would require a large amount of computer
time, and would not be very instructive.  We therefore arbitrarily
consider a binary to become ionized as soon as its apastron distance
exceeds 10\AU.

A collision between stars during the three-body encounter is assumed
to occur whenever two stars approach each other to within a distance
equal to the sum of their radii.  We thus neglect tidal effects, which
are relatively less important, given the competing three-body effects
that prevent subsequent orbital circularization after what would have
become a case of tidal capture if it had occurred in isolation.  We
also neglect mass loss during mergers: the mass of a merger remnant is
taken to be the sum of the masses of the stars involved.

To give a reasonable prescription for the radius of the remnant star
is a more difficult task, since the initial configuration of the
remnant, while in dynamical equilibrium, will be far from thermal
equilibrium, and therefore likely to extend well beyond its final
relaxed size.  For simplicity, we assign a merger product an initial
radius that is equal to the sum of the radii of the two (or three)
colliding stars, at least for the duration of the scattering
experiment.  This is a reasonable procedure, since the dynamical
encounter takes place over a time interval that is negligible compared
to both the thermal and the evolutionary time scales of the stars
involved.  As soon as the encounter is finished, we relax the
structure and the evolutionary state of the merger remnants to their
appropriate equilibrium configurations, according to the appropriate
stellar evolution model (see paper~I).

\subsection{Terminology}
\label{sec_term}
Here we present a list of the main terms used in the present paper.

\begin{itemize}
\item[]
{\it Blue straggler:} a main sequence star with a mass larger than the
	turn-off mass. \\ 
\item[]
{\it Coalescence:} the formation of a single
	object due to an unstable phase of mass transfer in a
	binary. \\ 
\item[]
{\it Collision:} result of a violent dynamical encounter
	between two or tree stars, leading to the formation of a
	single star. \\ 
\item[]
{\it Exchange:} a three-body encounter in which the incoming star
	replaces one of the binary components. \\ 
\item[]
{\it Ionization:} an event in which all stars involved emerge unbound.
	This can be the result of a supernova explosion, in the case
	of a two-body system, or it can occur as the result of a
	three-body encounter.  It can also occur when a binary has become
	progressively wider, and finally reaches an apastron value exceeding
	10\AU. \\
\item[]
{\it Merger:} the outcome of a process involving two or three stars
	which either collide or coalesce.  \\
\item[]
{\it Yellow straggler:} a (sub)giant with a mass
	larger than that of ordinary (sub)giants. \\ 
\item[]
{\it Resonance:} an intermediate state of a three-body encounter,
	in which all three stars are temporarily bound, and
	therefore undergo more than one close passage
	during the scattering process. \\ 
\item[]
{\it Preservation:} a three-body encounter in which the identity
	of the binary members remains unchanged. \\ 

\end{itemize}

\begin{table*}
% table 2
\caption[]{Parameters of the different model computations, and some
resulting numbers. Columns specify the model, the core radius, the
3-dimensional velocity dispersion for a $1\msun$ star, the time at
which encounters are started, the storage time interval, the stellar
number density in the core, and the fraction of the true core that we
simulate (for the computations with single stars only).  
For single stars we give the number of encounters per star
[$\st^{-1}$] and the mean time between encounters (see also paper~I);
for binaries we list the average number of encounters per binary
[$\st\st^{-1}$] and the mean time between encounters.  For tidal
binaries the mean time $\delta t_{tide}$ between capture events is
also given.
}
\begin{center}
\begin{tabular}{lcrrrrl|ll|lr|rrr} \hline
&&&&& & 
       &\multicolumn{2}{c}{single stars} 
       &\multicolumn{2}{c}{primordial} 
       & \multicolumn{3}{c}{tidal capture} \\
{\small Model} &$r_{\rm c}$&\vsun&$t_{\rm cc}$&$\Delta T$
      &$\log n$ & $f_c$
      & $n_{\rm enc}$ & $\tau_{\rm enc}$
      & $n_{\rm enc}$ & $\tau_{\rm enc}$
      & $\delta t_{\rm tide}$ & $n_{\rm enc}$ & $\tau_{\rm enc}$ \\ \hline
      &{\small [pc]} &{\small [km/s]}&{\small [Gyr]}&{\small [Myr]}
            &{\small [$\star$~pc$^{-3}$]} &
      &[$\st^{-1}$] &{\small [Myr]} 
      &[$\st\st^{-1}$] &{\small [Myr]} 
      &{\small [Myr]}&[$\st\st^{-1}$]&{\small [Myr]} \\  \hline
$S$ &4.0 &17.3& 0&10&3.92&0.298 &0.002&7.56 &6.7&2300 &14.0& & \\
$C$ &0.1 &17.3&10& 5&6.64&8.750 &0.258&5.81 &33.6&26.6 
				      &9.28 &11.5& 80.9  \\
\end{tabular}
\end{center}
\label{Tab_init}\end{table*}

\section{Results for the high-density system}
\label{sec_73}

Because the effect of encounters is strongest in a dense cluster core,
we first discuss the evolution of a population of primordial binaries
in a stellar system with a relatively high number density of stars,
and compare the results of this model with an identical population of
primordial binaries that did not experience any encounters.  The mass
function for this high-density cluster core (model $C$, for
``collapsed core'') is affected by mass segregation, causing it to be
relatively flat.  Therefore, for model $C$ we use equal numbers of
stars per unit mass interval between 0.1~\msun\ and the maximum mass
of a single star.  For binaries, primary and secondary masses,
semi-major axes and eccentricities are initialized as described in
sect.~\ref{sec_prim} (see also Table~\ref{Tab_initb}).

In model $C$, dynamical encounters start at $t_{\rm cc}=10$~Gyr and
the calculation is stopped at $16\,$Gyr.  Single stars, forming the
environment in which the binaries are evolved, were binned (in number
density and relative density of stellar types, and masses and radii of
the single stars) and stored at intervals of $\Delta t = 5$~Myr (see
Table~\ref{Tab_init} and sect.~\ref{sec_ssenv}).

In sect.~\ref{sec_nondyn} we discuss the results of computations in
which binaries evolve in isolation, without dynamical encounters.  The
evolution of primordial binaries that are allowed to encounter single
stars are presented in sect.~\ref{sec_pbin}.  The results for
tidal binaries are discussed in sect.~\ref{sec_tidalC}.

\subsection{Primordial binaries without encounters}
\label{sec_nondyn}

For the computation of primordial binaries without any encounters, a
total of 50000 binaries are followed.  It is convenient to define the
fractional lifetime of a possible binary state [(ms, ms), (ms, gs),
(gs, gs), etc., where the terms are defined in the caption of
Table~\ref{Tab_blife}] as the total time that all binaries spend in that
state divided by the total lifetime of all binaries.  The average
lifetime of a non-dynamically evolving binary is about 4.88~Gyr (from
$t=10$ to 16~Gyr, see Table~\ref{Tab_blife}).  Binaries that contain
two main-sequence stars are, with a fractional lifetime of 90.0\%,
most common.  On average, a binary spends 7.0\% of its time as a
main-sequence star with a white dwarf companion and 1.5\% of
the time with a (sub)giant.  White-dwarf--white-dwarf binaries have a
fractional lifetime of 1.3\%.  The remaining $\sim 0.2\%$ corresponds
mainly to a (sub)giant accompanied by a white dwarf (see
Table~\ref{Tab_blife}).

Table~\ref{Tab_merge} gives the fraction of binaries that merge after
an unstable phase of mass transfer.  Overall, 9.3\% of all binaries
merge during the period of the calculation.  Most mergers (82.2\%) are
the result of an unstable phase of mass transfer from a (sub)giant
onto a main-sequence star.  A considerable fraction of binaries
survive this first phase of mass transfer, only to merge as soon as
the main-sequence star in turn fills its Roche lobe and transfers mass
onto the white dwarf.  A small number of mergers (120 in total)
originate from white-dwarf pairs ($50000 \times 0.093 \times 0.026$,
see Table~\ref{Tab_merge}, second column), closely followed by merging
main-sequence binaries (112 in total).

\begin{figure}
\hspace*{0.5cm}
%\epsfxsize = 4.5cm
%\hspace*{1.cm}
\psfig{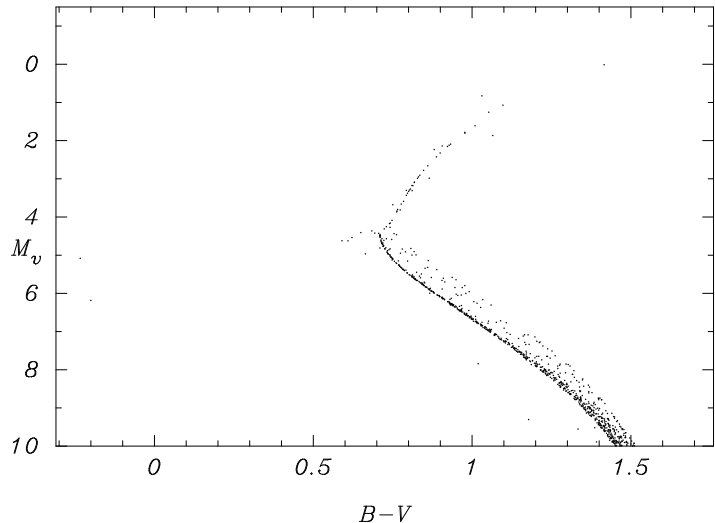}
%\epsffile{fig_hrdCn.ps}
\caption[]{Hertzsprung-Russell diagram of a population of primordial
binaries in model $C$, with no interactions with other cluster
members, at a system age of 12~Gyr.  Of the 5000 binaries we started
with to produce this plot, 7.9\% have coalesced in an unstable phase
of mass transfer; the remainder survived.  The axes are chosen to
match the Hertzsprung-Russell diagram with encounters between single
stars presented in paper~I (see also \cite{pz96}).  The main-sequence
and the (sub)giant branch are clearly visible, as is the second
main-sequence extending to $\sim$0.75 magnitudes above the zero-age
main-sequence.  White dwarf binaries lie to the left with a color
index of $B-V \la 0$.  The few blue stragglers arise from internal
binary evolution (stable mass transfer).}
\label{fig_hrdCn}\end{figure}

Figure~\ref{fig_hrdCn} shows the Hertzsprung-Russell diagram at 12 Gyr
of a population of 5000 primordial binaries in model C.  (A fraction
of the total number of binaries computed is chosen to facilitate
comparison with the Hertzsprung-Russell diagrams of the other models.)
The binaries at $B-V\sim 1.1$ -- 1.4 and $M_v \sim 9$ -- 10 (below the
main sequence), contain a white dwarf and a low mass main-sequence
star.  In the absence of dynamical encounters, only a small number of
blue stragglers are formed in the population of primordial binaries
 from model $C$ (see also Fig.~\ref{fig_hrdCn}).  The majority of blue
stragglers formed by binary evolution are the result of an unstable
phase of mass transfer leading to coalescence; these single stars are
not presented in Fig.~\ref{fig_hrdCn}; only blue stragglers formed by
stable mass transfer appear on the diagram.  From
Table~\ref{Tab_merge} we see that the majority of mergers occur
between a main-sequence star and a (sub)giant, which does not lead to
the formation of a blue straggler.

\begin{figure}
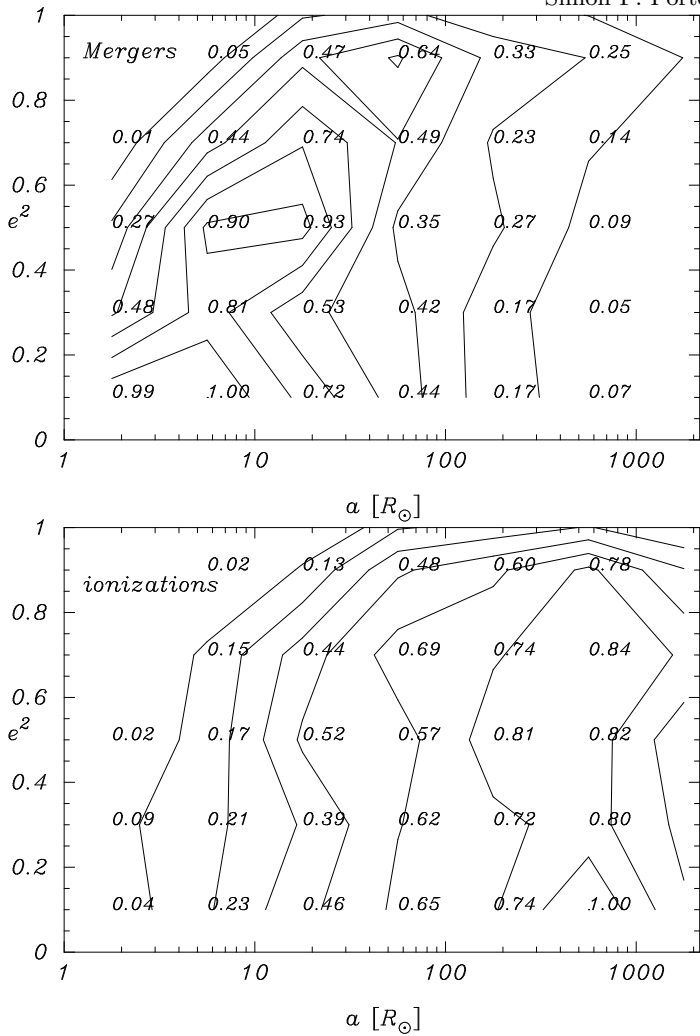

\hspace*{0.5cm}
\epsfxsize = 4.5cm
%\epsffile{fig_init_collC.ps}
\psfig{figure=fig_init_collC.ps,width=4.5cm}

\vspace*{0.75cm}
\hspace*{0.5cm}
\epsfxsize = 4.5cm
%\epsffile{fig_init_ionC.ps}
\psfig{figure=fig_init_ionC.ps,width=4.5cm}
\caption[]{In the $\log a$ -- $e^2$ plane, equal numbers of initial
binaries are selected per unit area, with the exception of the upper
left corner, where the periastron distance is unacceptably small.  The
numbers indicate for each location in this plane the fraction of
primordial binaries in model $C$ that merge (top figure) or are
ionized (bottom figure) relative to the maximum in that plane.  The
contours, at 12.5\% intervals, follow the absolute number of events.
The majority of binaries that end their lifetime as merged objects are
found in the lower left corner of the semi-major axis -- eccentricity
plane, indicating that, as expected, binaries with short orbital
periods and small eccentricities are more likely to merge, while
binaries with initially large orbital separations are more likely to
be ionized. 
}
\label{fig_coll_ionC}\end{figure}

\begin{table}
%\vspace*{-2cm}
\caption[]{Average lifetime of binaries in our models.
The first row gives the average lifetime of all binaries (in Gyr).
Subsequent rows give the fraction of this lifetime that an average
binary spends in a particular phase.  For example, a primordial binary
in model $C$ spends 78.0\%\ (693~Myr) of its time as a binary with two
main-sequence stars (ms, ms).  Abbreviations are as follows: ms:
main-sequence star; gs: giant; wd: white dwarf; ns: neutron star; bh:
black hole; we use \st\ to mean any star, regardless of type.  
Note that we do not list Thorne-\Zytkow objects, helium stars
and Wolf-Rayet stars, nor do we discriminate between various
evolutionary stages of (sub)giants, horizontal-branch stars and
super giants.}
\begin{center}
\begin{tabular}{l|lll|ll} \hline
binary    & \multicolumn{3}{c}{Model $C$}   & \multicolumn{2}{c}{Model $S$} \\
          &non-d&prim  &tide  &non-d&prim \\ \hline
(\st, \st)&4.88 &0.892 &0.928 &15.58 &15.62   \\
(\st, \st)&1    &1     &1     &1    &1          \\ \hline
(ms, ms)  &0.900&0.780 &0.241 &0.988&0.961   \\
(ms, gs)  &0.015&0.004 &0.007 &0.002&0.003   \\
(ms, wd)  &0.070&0.125 &0.454 &0.007&0.028   \\
(ms, ns)  &0.000&0.035 &0.129 &0.000&0.003 \\
(ms, bh)  &0.000&0.008 &0.026 &0.000&0.003  \\
(gs, gs)  &0.000&0.000 &0.002 &0.000&0.000  \\
(gs, wd)  &0.002&0.006 &0.026 &0.000&0.000   \\
(gs, ns/bh)&0.000&0.001 &0.009 &0.000&0.000  \\
(wd, wd)  &0.013&0.017 &0.030 &0.002&0.001   \\
(wd, ns)  &0.000&0.011 &0.022 &0.000&0.000  \\
(wd, bh)  &0.000&0.006 &0.017 &0.000&0.000  \\
(ns, ns)  &0.000&0.004 &0.005 &0.000&0.000  \\ 
(ns, bh)  &0.000&0.003 &0.007 &0.000&0.000  \\ \hline
\end{tabular}
\end{center}
\label{Tab_blife}\end{table}

\begin{figure}
\hspace*{0.5cm}
%\epsfxsize = 4.5cm
%\hspace*{1.cm}
\psfig{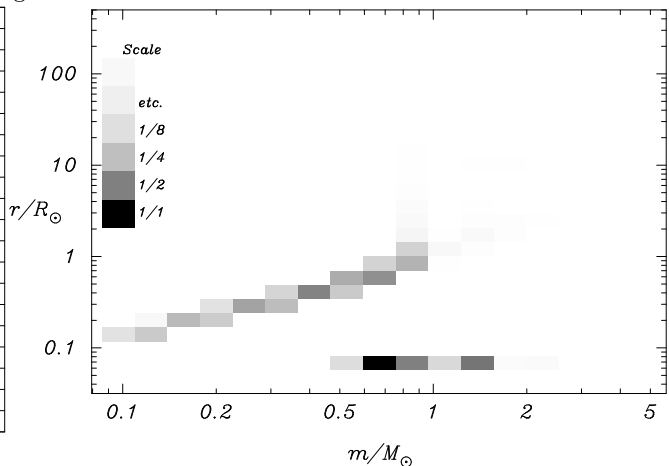}
%\epsffile{fig_rencfC.ps}
\caption[]{Relative encounter probabilities for a circular binary with
total mass $1\msun$ and semi-major axis $1\AU$, as a function of the
mass and radius of the other star involved in the encounter, for model
calculation $C$ at time $t=12\,$Gyr, when the turn-off mass is $M_{\rm
to}=0.91\,\msun$.  Darker shades indicate higher probabilities.
Compact stars (nominally with zero radius) are shown as a bar below
0.1$\,\rsun$: neutron stars lie between 1.34 and 2 $\msun$, white
dwarfs and black holes lie at lower and higher masses respectively.
All other stars with mass above the turn-off mass are products of
previous encounters.  The vertical bar at the upper left corner
indicates the scaling.  The lowest square corresponds to an encounter
rate of one per 193~Myr, decreasing by a factor of two for each subsequent
square.  The integrated encounter frequency of the $1\msun, 1\AU$
binary is 1 encounter every 30.3~Myr.  About 16\%\ of the encounters
occur with a $\sim 0.7 \msun$ white dwarf (black square).
}
\label{fig_rencfC}\end{figure}

\subsection{Primordial binaries with stellar encounters}
\label{sec_pbin}

A total of 5000 primordial binaries were initialized (see
Table~\ref{Tab_initb}) and evolved in the dynamical stellar
environment of single stars.  The results of the scattering
experiments performed on these binaries are summarized in
Table~\ref{Tab_scatter}.

\subsubsection{Encounter outcomes}

If a binary survives an encounter with another cluster member, its
orbital elements, and possibly also its stellar components, are
affected.  Stellar evolution also affects these binary parameters.
Both competing physical effects are computed simultaneously in order
to follow the evolution of the binary population.  The computation is
terminated when a binary is dissociated, its components merge (without
leaving the binary) or it escapes from the potential well of the stellar
system after an encounter.  Since a single binary can undergo multiple
encounters, the total number of encounters is much larger than the
initial number of binaries (see Table~\ref{Tab_init}).

Table~\ref{Tab_term} presents the probabilities of the various binary
termination channels.  In the computation of the primordial binaries
in model $C$, ionization is the dominant termination channel (see
Table~\ref{Tab_term}).  A colliding encounter often results in the
formation of a yellow straggler in a relatively wide binary.  The size
of the yellow straggler makes subsequent encounters a dangerous
process, however, and a collision or an unstable phase of mass
transfer can easily destroy the binary.  The flat mass function allows
many relatively massive, low-velocity stars to enter resonance
encounters, and a considerable fraction of the primordial binaries
collide or coalesce.

\begin{table*}
\caption[]{Relative importance of various types of stellar
collisions during dynamical encounters or mergers following an
unstable phase of mass transfer.  For each model we give the sum of
the collision and coalescence rates (the total is normalized to unity)
and the coalescence fraction.  For non-dynamically evolving binaries
(non-d), the coalescence rate is the total rate.  Due to their small
rates Thorne-\Zytkow objects are not included, and triple collisions
\{\st, \st, \st\} and mergers between two remnants \{ns/bh, ns/bh\} are not
further specified.  In some of the models the initial conditions do
not allow certain stellar types to merge; this is indicated with a
dash ``--''.  The total number of collisions per binary can be larger
than unity (as is the case for the tidal binaries in model
$C$), a new binary can be the result of a collision during a triple
encounter.
}
\begin{center}
\begin{tabular}{l|llrlr|llrr} \hline
          &            & \multicolumn{4}{c}{Model $C$} &
           &              \multicolumn{3}{c}{Model $S$} \\
origin     & non-d& \multicolumn{2}{c}{primordial}
                        & \multicolumn{2}{c}{tidal}
           &  non-d
                        & \multicolumn{2}{c}{primordial}
                        & tidal \\
&merged&merged&coal.&merged&coal.&merged&merged&coal.&coal. \\ \hline
per binary &0.093  &0.571&25\%&1.032&39.6\%&0.094&0.116&16\%&0.53 \\
$n_{\rm merge}$&1  &1    &[\%]&1    &[\%]&1    &1    &[\%]&1    \\
\{ms, ms\} &0.024  &0.441&21 &0.327&66   &0.36 &0.80 &19  &0.91 \\
\{ms, gs\} &0.822  &0.106&70 &0.072&54   &0.39 &0.06 &71  &0.05  \\
\{ms, wd\} &0.115  &0.187&15 &0.236&21   &0.08 &0.11 &48  &0.05 \\
\{ms, ns\} &0.000  &0.081& 0 &0.087& 1   &0.00 &0.01 &0  &0 \\
\{ms, bh\} &0.000  &0.009& 1 &0.017& 2   &0.00 &0.00 &0  & 0   \\
\{gs, gs\} &0.007  &0.002&20 &0.005&61   &0.01 &0.00 &0  &0 \\
\{gs, wd\} &0.006  &0.023&26 &0.047&32   &0.03 &0.00 &0  &0 \\
\{gs, ns/bh\}&0.000&0.015& 3 &0.027&11   &0.00 &0.00 & 0  &0  \\
\{wd, wd\} &0.026  &0.046&70 &0.055&77   &0.13 &0.02 &100 &0.1 \\
\{wd, ns/bh\}&0.000&0.015&60 &0.029&71   &0.00 &0.00 &0   & 0   \\
\{ns/bh, 
ns/bh\}      &0.000&0.004&70 &0.004&85   &0.00 &0.00 &100 &0 \\
\{\st,\st,\st\}&-- &0.071&-- &0.104&--   & --  &0.01 & -- &0 \\ \hline
\end{tabular}
\end{center}
\label{Tab_merge}\end{table*}

Figure~\ref{fig_coll_ionC} shows the number of binaries, as a function
of semi-major axis and eccentricity, that end their lifetime in a
single object (upper panel), or that are ionized due to a dynamical
encounter or by the dispersion criterion (lower panel).  The initial
conditions for all binaries are sampled homogeneously over the entire
plane, with the exception of the upper left (small period and large
eccentricity) corner.  As expected, mergers occur more frequently in
binaries with initially small orbital separations, while binaries with
larger initial semi-major axes tend to be ionized.  Near $a_{\rm max}$
the ionization fraction approaches 100\%, indicating that our choice
of maximum semi-major axis was large enough.

The total fraction of encounters leading to preservation is 95\% (see
Table~\ref{Tab_scatter}).  Only 11 non-preservation encounters
resulted from encounters with an impact parameter exceededing 90\% of
the maximum impact parameter, confirming that our choice for the
maximum impact parameter was large enough to ensure that not too many
`interesting' encounters are missed, and small enough to get
reasonable statistics on unlikely encounters.

Only a small fraction ($\sim 1.8$\%) of the preservations result from
a resonant encounter (indicated in the third column of
Table~\ref{Tab_scatter}).  Interactions where the primary is exchanged
for the incoming field star [indicated as 1, (2, 3) in
Table~\ref{Tab_scatter}] are less than half as frequent as
interactions where the incoming star takes the place of the secondary:
2, (1, 3).  More than half of the secondary exchange encounters occur
in resonances, whereas the majority ($\sim 70$\%) of exchanged
primaries are exchanged in prompt encounters.  Note that only a small
fraction of all encounters result in ionization.  This is not surprising,
since such an encounter can occur at most once per binary, while other
interaction outcomes can occur many times (the same applies to
collision-escape and triple collisions, to be discussed in the next
paragraph).

Since the total number of collisions is rather small we normalize them
separately to unity (see the lower part of Table~\ref{Tab_scatter}).
We subdivide the collisional encounters into (1) a collision
(indicated by braces) and the subsequent formation of a binary 
(indicated by parentheses) (2) ionization (no parentheses) of the
binary after the collision event and (3) the collision of all three
stars, leaving a single object (braces).  The collision--binary and
collision--ionization events can both be further separated into three
subclasses, depending on which two stars merge (see
Table~\ref{Tab_scatter}).  Only a small fraction of triple mergers
occur after a resonance encounter.  This can be understood when one
realizes that the incoming single star might well be comparable in
size to the binary; a giant that encounters a close binary will
collide with both binary components before having any chance to enter
a resonant encounter.

\subsubsection{Comparison with two-body collisions}

\begin{table}
%\vspace*{-2cm}
\caption[]{The results of dynamical encounters.  For each possible
outcome of a single scattering experiment (first column) the table
gives encounter rates and the fraction (as a percentage) of encounters
that were resonances.  Data are presented for both primordial and
tidal binaries, for high-density model $C$, and Salpeter model $S$.
The total number of encounters $n_{\rm enc}$ for each model is
normalized to 100.  Hierarchical resonances are rare, and are combined
with democratic resonances.  The lower part of the table further
subdivides the merging encounters, where for each model the total
number of collisions is normalized to unity.  Parentheses indicate a
bound system, braces indicate a collision.  A dash ``--'' indicates that
this type of encounter is not possible.  The total number of resonance
preservation encounters per binary in model $C$ is 0.57 ($0.950 \times
1.8\%\,n_{\rm enc} = 1.7\%\, n_{\rm enc}$; $n_{\rm enc}=33.6$, from
Table~\ref{Tab_init} with 5000 initial binaries).  }
\begin{center}
\begin{tabular}{l|rrrr|rrr} \hline
      & \multicolumn{4}{c}{Model $C$} & \multicolumn{3}{c}{Model $S$} \\
      & \multicolumn{2}{c}{primordial} & \multicolumn{2}{c}{tidal}
      & \multicolumn{2}{c}{primordial} & tidal \\
      & total&res. & total&res.&tot.&res.& tot. \\
$n_{\rm enc}$     &100\mbox{~ ~} &[\%]&100\mbox{~ ~}
                  &[\%]&100\mbox{~ ~}&[\%]&100\mbox{~ ~}\\ \hline
(1, 2), 3         &95.0 & 1.8 &91.0&7.0  &96.9& 4 &99.0 \\
1, (2, 3)         &0.9  &30.8 & 0.8&58.1 & 0.6&51 & 0.0 \\
2, (1, 3)         &2.0  &57.0 & 2.7&71.4 & 1.0&63 & 0.2 \\
1, 2, 3           &0.8  & --  & 0.1& --  &   0&-- & 0.0 \\
merged            &1.3  &47.4 & 5.4&44.7 & 0.1&27 &0.8 \\ \hline
merged:           & 1\mbox{~ ~ ~}&    &1\mbox{~ ~ ~} &    
                  &1\mbox{~ ~}&   &1\mbox{~ ~}      \\
(1, \{2,3\})      &0.19  &67  &0.21 &48  &0.3&26 &0.1 \\
(2, \{1,3\})      &0.18  &61  &0.19 &49  &0.4&51 &0.5 \\
(3, \{1,2\})      &0.26  &68  &0.26 &74  &0.3&81 &0.1 \\
 1, \{2,3\}       &0.02  &21  &0.02 &33  &0.0&55 &0.0    \\
 2, \{1,3\}       &0.02  &41  &0.02 &55  &0.0&37 &0.0 \\
 3, \{1,2\}       &0.21  &12  &0.11 &27  &0.1&33 &0.1 \\
\{1, 2, 3\}       &0.10  & 9  &0.17 &10  &0.0&50 &0.0   \\ \hline
\end{tabular}
\end{center}
\label{Tab_scatter}\end{table}

\begin{table}
\caption[]{Probability of various outcomes of a binary
in the different model computations.  The meanings of the terms used
are explained in sect.~\ref{sec_term}.}
\begin{center}
\begin{tabular}{l|rlr} \hline
origin      & \multicolumn{2}{c}{Model $C$}
            & Model $S$ \\
            & prim & tidal&prim      \\ \hline
preservation&0.144 &0.215 &0.955     \\
ionization  &0.599 &0.110 &0.017     \\
collision   &0.149 &0.198 &0.014     \\
coalescence  &0.058 &0.387 &0.018     \\
escape      &0.050 &0.090 &0.000     \\ \hline
\end{tabular}
\end{center}
\label{Tab_term}\end{table}

Figure~\ref{fig_rencfC} depicts the relative probability for a circular
binary, with total mass 1~\msun and semi-major axis 1~\AU, to
encounter a single star in the stellar system.  Shaded squares
indicate the binning used to improve performance, darker shades
indicating higher encounter probabilities.  A comparison between
Fig.~\ref{fig_rencfC} and Fig.~3 of paper~I reveal the effect of the
larger size of the binary on the relative encounter probabilities: the
binary encounter probabilities follow the relative densities of the
various stellar species, whereas the probability distribution for the
1~\msun single star from paper~I is more strongly affected by a few
individual stars with large interaction cross sections.  Collision
products formed during the previous evolution of the stellar system
are present and can experience encounters with the binary.  Blue and
yellow stragglers are evident in Fig.~\ref{fig_rencfC} as stars with
masses larger than {$\sim1$~\msun} and with radii smaller and larger,
respectively, than $\sim 2$~\rsun.

\begin{figure*}
\vspace*{-5.5cm}
\hspace*{5.0cm}
\epsfxsize = 2.0cm
%\epsffile{fig_rcollfCCT.ps}
\psfig{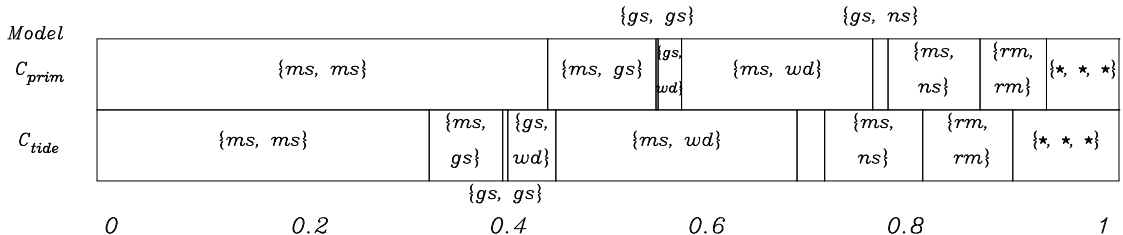}
\caption[]{Relative merger frequencies (normalized to the total number
of mergers) between various stellar species, from dynamical model $C$.
The upper bar shows results for primordial binaries, the lower for
tidal binaries. 
The abbreviation rm (for remnant) is used for both neutron stars and
black holes.  
}
\label{fig_collCCT}\end{figure*}

Figure~\ref{fig_collCCT} visualizes the relative merger frequencies for
the dynamically evolving model $C$.  Comparison with Fig.~4 from
paper~I, which gives the relative collision frequencies between
different stellar species in model $C$, shows a enhancement of the
number of mergers between white dwarfs and other stars; other merger
frequencies are suppressed. The small radii of white dwarfs mean that
they are not favored in encounters between single stars. A white dwarf
is considerably more likely to merge with another star during a binary
interaction.

In paper~I the only channel through which a blue straggler could form
was a collision between two single stars.  Although more than 20\% of
the stars in the high-density stellar system experienced a collision,
the number of blue stragglers visible at any instant did not exceed
about 3\% (per star).  The presence of primordial binaries provides an
efficient channel of up to about 1\% per binary for the formation of
blue stragglers, either single or in binaries (see Fig.~\ref{fig_nbss}).

\subsubsection{Comparison with binaries without encounters}

\begin{figure}
\hspace*{0.5cm}
\epsfxsize = 4.5cm
%\epsffile{fig_hrdC.ps}
\psfig{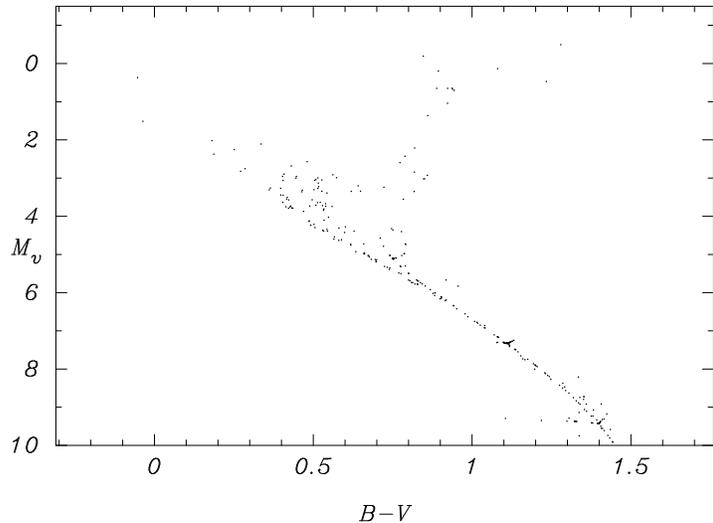}
\caption[]{Hertzsprung-Russell diagram of a population of primordial
binaries at a system age of 12~Gyr. The dots indicate binaries with
visual magnitude smaller than 10 that survived the first 2~Gyr of
dynamical evolution in model $C$. The scale is chosen to match the
Hertzsprung-Russell diagram with encounters between single stars from
paper~I. The small number of binaries reflects their short lifetime.
At this time, the main-sequence turn-off is around $M_v=4.5$, with
$B-V \approx 0.7$ (effective temperature of 5500~K).}
\label{fig_hrdC}\end{figure}

Figure~\ref{fig_hrdC} shows the Hertzsprung-Russell diagram at 12 Gyr of
the primordial binaries in model C with encounters. About 41\% 
binaries survive the first 2~Gyr of dynamical evolution in the host
stellar core.  The objects at $B-V\sim 1.3$ and $M_v \sim 9$ -- 10
(below the main sequence) are binaries comprising a low-mass
main-sequence star and a white dwarf.  There are very few giants among the
primordial binaries---the majority of the
binaries that are wide enough to contain a
(sub)giant were destroyed in the first few hundred million years of
the dynamical evolution of the stellar system.  The large number of
binaries still present at $t=12$~Gyr in Fig.~\ref{fig_hrdCn} compared
to the small number of binaries in Fig.~\ref{fig_hrdC} reveals that
the primordial binaries that experience occasional encounters with
other cluster members have a much shorter lifetime then the
non-dynamically evolving binaries (see also Table~\ref{Tab_blife},
first row).

A ``gap'' between the turn-off of the stellar system and the more
massive blue stragglers is visible in Fig.~\ref{fig_hrdC}.  This gap
is not as obvious as the gap in Fig.~5 of paper~I, but its origin is
similar: The majority of all blue stragglers formed are the product of
collisions.  Because they tend to form from low-mass, relatively
unevolved stars, and because their evolution time scales are long,
blue stragglers just above the turn-off still lie close to the
zero-age main sequence, as can be seen in Fig.~\ref{fig_hrdC}.  More
massive blue stragglers evolve faster, and consequently have a broader
distribution, extending from the zero-age main-sequence to the
terminal-age main-sequence (see also \cite{pz96}).  In our models, the
gap is largely the result of the dynamics (and hence most
blue-straggler formation) being switched on at a specific time
($t_{cc}$).  It is unclear to what extent this feature will persist in
models with a more sophisticated treatment of the cluster dynamics.

Binaries evolving in a dynamical environment show a rich diversity
compared to binaries that evolve non-dynamically (see
Table~\ref{Tab_blife}).  The formation of a binary containing a
neutron star or a black hole is very rare in a non-dynamically
evolving stellar system.  However, the relatively large masses and
small radii of these stellar evolution remnants make them excellent
potential exchange partners in dynamical encounters.  Once a massive
remnant is exchanged into a binary, it is likely to be the more
massive component, and not easily exchanged for another, lower mass,
star (see Table~\ref{Tab_scatter}).  This results in a strong
overabundance of binaries that contain at least one white dwarf,
neutron star or black hole.  An exchange interaction with a (sub)giant
is not likely---the large radius of the giant generally leads to a
collision with one or other of the stars during the encounter.

Table~\ref{Tab_merge} illustrates how dynamical encounters
dramatically enhance the probability of mergers (collisions as well as
coalescence), and open the possibility of triple collisions.  Although
most merger rates are increased by dynamical encounters, the total
number of \{ms, gs\} coalescence actually decreases considerably:
4.2\% ($0.571 \times 0.106 \times 0.70$) versus 7.6\% ($0.093 \times
0.822$) for the non-dynamically evolving case.  The reason is that
binaries with orbital periods large enough to contain a (sub)giant
have large cross sections for dynamical encounters, and tend to be
destroyed before the primary has had time to ascend the giant branch.

\begin{figure}
\hspace*{0.5cm}
\epsfxsize = 4.5cm
%\epsffile{fig_binsurv.ps}
\psfig{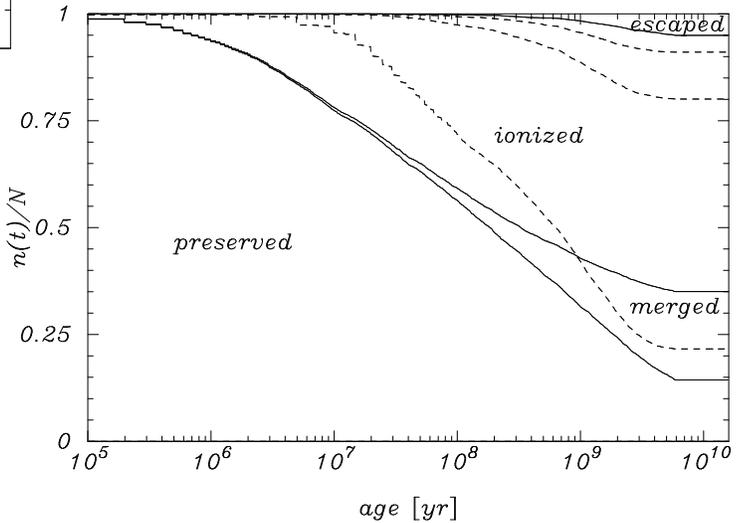}
\caption[]{Fate of binaries in Model $C$ as a function of time elapsed
since the onset of dynamical encounters at 10 Gyr.  The solid lines
separate the possible outcomes for primordial binaries: preservation
(i.e. binaries which either had no encounters or still are a binary
following an encounter), merger (after Roche overflow or following a
collision), ionization (disrupted by a supernova, dispersed because
the separation at apocenter exceeds 10\AU\, 
or ionized by an encounter), and escape.  For example,
after $2 \times 10^8\,$yr of encounters, $\sim 50$\%\ of the primordial
binaries remain, $\sim 5$\%\ have merged, and $\sim 45$\%\ have been
ionized; $<1$\%\ have escaped.  The dashed lines indicate the same
data for tidal binaries, where now ``age'' refers to time
since the binary formed.  As expected, the average lifetime of
primordial binaries is much shorter than that of tidally formed
binaries.
}
\label{fig_binsurv}\end{figure}

Figure~\ref{fig_binsurv} presents the overall evolution of the binary
population, as a function of time since 10 Gyr (time $t_{cc}$, when
the dynamical encounters were turned on).  Note that half of the
primordial binaries are destroyed within $\sim 100$~Myr.  

\subsubsection{Blue straggler evolution}

We identify a blue straggler as a main-sequence star with a mass
larger than the turn-off.  In real clusters, it is generally hard to
discriminate between main-sequence stars close to the turn-off and
actual blue stragglers (also the identification of blue stragglers
within the blue horizontal branch can be difficult in some cases).  In
our models, blue stragglers can be formed via number of different
channels, leading to several distinct classes of stragglers---a single
blue straggler can be formed by an unstable phase of mass transfer in
a binary, or by a collision in a 2- or 3-body encounter.  From
Table~\ref{Tab_merge}, we see that most blue stragglers formed in
model C are the result of dynamical encounters.

If a blue straggler is formed via a triple collision in a
binary--single-star encounter, its mass will likely be considerably
larger than that of most other blue stragglers in the stellar system.
The lifetime of such high-mass blue stragglers (masses more than twice
the turn-off mass, say) are very short, so their number at any moment
is too small to contribute significantly to the overall population.
The properties of blue stragglers formed by 2-body collisions between
single stars are discussed in Paper~I; the overall mass distribution
of such blue stragglers is not significantly different when the
collisions are mediated by primordial binaries.  

Blue stragglers formed by unstable mass transfer generally have
relatively small masses.  Such a merger, and the subsequent formation
of a blue straggler, can only occur via type $A$ mass transfer
(\cite{kw67}) when both stars are still processing hydrogen, 
which is unstable only if the mass ratio of the binary
is small.  The blue straggler thus formed is consequently slightly
only more massive than the turn-off.  Stable mass transfer leads to
the formation of a considerably more massive blue straggler in a
circular orbit around a Roche-lobe filling main-sequence star or a
remnant from the mass-transfer phase (a helium star, white dwarf, or
possibly a neutron star or a black hole).  In our models, binaries
containing a blue straggler and a mass-transfer remnant are often
found to have a small eccentricity, induced by subsequent encounters
with other stars.  In contrast, a binary blue straggler formed by a
collision during a 3-body encounter is generally highly eccentric, and
the companion is not limited to a mass-transfer remnant or a
Roche-lobe-filling star, but can be any type of star.

\begin{figure}
\hspace*{0.5cm}
\epsfxsize = 4.5cm
%\epsffile{n_byssCIIT.ps}
\psfig{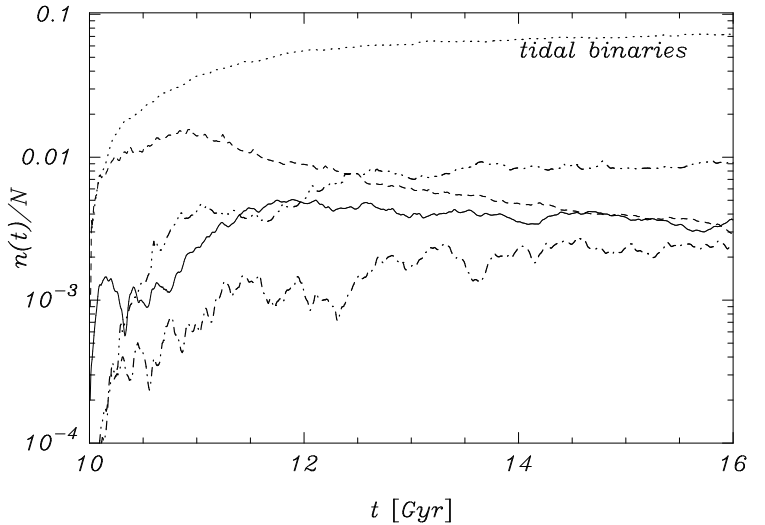}
\caption[]{Number of blue and yellow stragglers
formed in model $C$, with primordial binaries (dashed and solid lines,
respectively) and with tidal binaries (dash-dotted and
dash-dot-dot-dotted lines, respectively), as functions of
time.  The total number of tidal binaries is also shown
(dotted line).  All numbers are given as fractions of the total number
of binaries modeled in the computations.}
\label{fig_nbss}\end{figure}

Figure~\ref{fig_nbss} shows the numbers of blue and yellow stragglers in
model C, as functions of time.  During the first few million years of
evolution in the high-density stellar system the numbers of blue and
yellow stragglers rise steeply, eventually reaching equilibrium
between formation and destruction of both types of objects.  The
fraction of yellow stragglers arrives at its maximum about 180~Myr
after the onset of dynamical encounters.  By this stage more than half
of the original number of binaries have been destroyed, and the
production rate of yellow stragglers decreases rapidly.  At $t \approx
10.7$~Gyr only $\sim 35$\% of the primordial binaries are still
present in the stellar system (see Fig.~\ref{fig_binsurv}), and the
formation rate of blue stragglers becomes comparable to their
destruction rate; the number of blue stragglers then starts to
decrease and continues to do so until the end of the simulation.
As blue stragglers evolve into yellow stragglers, the number of the
latter increases again (from $t \approx 10.5$~Gyr) until the
transformation of blue stragglers into yellow stragglers and the
evolution of yellow stragglers into white dwarfs reaches equilibrium
(at around 11.5~Gyr).

\subsection{Tidal binaries in a dense cluster core}
\label{sec_tidalC}

The rate at which close flybys between stars result in the formation
of tidal binaries is only slightly smaller than the single-star
collision rate (see Table~\ref{Tab_init}).  This can be understood by
noting that the dominant term in the encounter cross section is linear
in the distance of closest approach (due to gravitational focusing,
see Eq.~\ref{eq:sigma}), and tidal capture is effective within a
distance of roughly 3--4 times the radius of the larger star involved.
Note that the collision rate between single stars (see
Table~\ref{Tab_init}) is smaller than computed in paper~I.  This is
caused mainly by differences in the maximum periastron distance
$d_{max}$ for which a physical collision can occur.  As already
discussed, in paper~I we adopted a value of $d_{max}$ equal to twice
the sum of the stellar radii, in an attempt to include (in an
approximate way) tidal captures leading immediately to merging into
the total collision rate.  In this paper we take $d_{max}$ for
collisions equal to the sum of the stellar radii, with a separate
treatment of tidal capture.  However, tidal binaries which begin
Roche-lobe overflow directly after formation are also counted as
collisions.

Capture between two remnants does not occur in our simulations, since
both stars have nominally zero radius.  The only way for a binary
neutron star to form is via an exchange interaction, where an incoming
neutron star replaces the companion of a neutron star already in a
binary.  If the exchanging binary has a short orbital period, the
recoil velocity produced in the exchange interaction is generally
large, and a considerable fraction of these interactions lead to
escape of the neutron-star--neutron-star binary.

The initial conditions of tidal binaries differ from those
of primordial binaries.  Tidal binaries can be formed (a) at any
moment between $t=t_{\rm cc}$ and 16~Gyr, (b) with arbitrary stellar
types and masses at the moment of formation, and (c) with very small
semi-major axes and eccentricities.  The formation rate of tidal
binaries is continuous from the instant when the stellar
density reaches the value required for two-body encounters to become
important until the end of the simulation.  After an initial startup
period of about 1~Gyr, the total number of tidal binaries remains
roughly constant (see Fig.~\ref{fig_nbss}).

Initially there are no black holes in the stellar system, not even in
the background single-star cluster: all black holes are formed from
neutron stars that accrete sufficient material after a collision with
another cluster member.  This accretion process takes some time (about
1~Gyr), and the first tidal capture of a black hole by a main-sequence
or giant star does not occur until $t=11.5$~Gyr.  Black-hole tidal
binaries are more common than black-hole primordial binaries (see
Table~\ref{Tab_blife}), mainly because the number of targets for
tidal capture is much greater than the number of surviving primordial
binaries available for exchange.

\begin{figure}
\hspace*{0.5cm}
\epsfxsize = 4.5cm
%\epsffile{fig_init_tc.ps}
\psfig{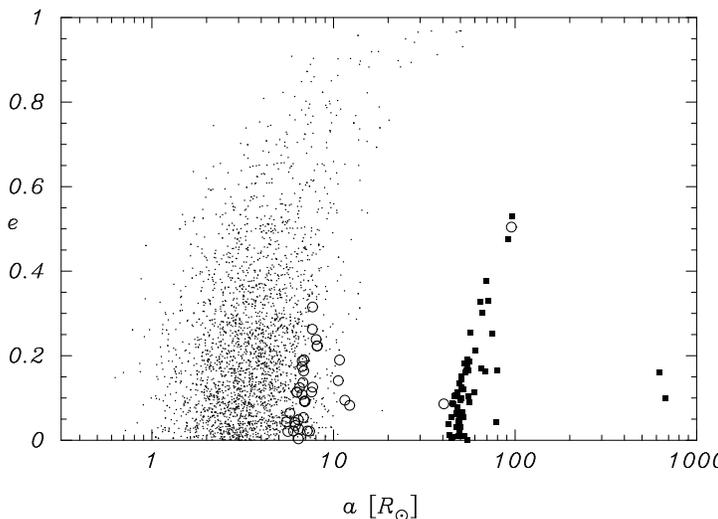}
\caption[]{The initial semi-major axes and eccentricities of 3000
tidal binaries with non-zero eccentricity, from model $C$.
Dots, circles and squares indicate binaries containing a white dwarf
and a main-sequence star, an early subgiant, and a horizontal branch star,
respectively.  The two squares at the right side of the figure are
binaries with a white dwarf and a super giant.
A subgiant spends most of its time at the bottom of the giant branch where
its radius is smaller than that of a horizontal-branch star, for this
reason the majority of the tidal binaries with a subgiant have a relatively
small semi-major axis. 
The majority of tidal binaries are born with
zero eccentricity and a short orbital period.  (This statement,
however, depends strongly on the algorithm adopted for their
formation.)}
\label{fig_init_tc}\end{figure}

The majority ($\sim 60$\%) of tidal binaries have zero eccentricity.
The distribution of the orbital parameters of the remainder, born with
finite eccentricity, is illustrated in Fig.~\ref{fig_init_tc}.
The various stellar types that
can capture a white dwarf populate different areas in the diagram.
For each stellar type that captures a white dwarf, there is a minimum
periastron distance for which successful (i.e. non-colliding) capture
can occur.  This is nicely demonstrated by the stars on the horizontal
branch, all which have about the same radius.  Note that this reflects
the peculiarities of the algorithm adopted for the formation of
tidal binaries; what nature does in these cases is unclear.

Because of their small orbital separations, relatively few tidal
binaries are involved in encounters with other stars, and
they are unlikely to become ionized when they do interact.  Compared
to primordial binaries, tidal binaries have long time
intervals between encounters, and do not get ionized or exchanged so easily
(see Table~\ref{Tab_term}).  
Not surprisingly, those
encounters that do occur have a larger fraction of resonance
encounters, in which collisions are frequent (see
Table~\ref{Tab_scatter} and Table~\ref{Tab_merge}).  The majority of
tidal binaries are destroyed by coalescence, collision or ejection
 from the stellar system (see Table~\ref{Tab_term} and
Fig.~\ref{fig_binsurv}).  The half lifetime for tidal binaries (lower
dashed line in Fig.~\ref{fig_binsurv}) is more than twice as large as
for the primordial binaries.

Due to their small encounter rates, tidal binaries are more strongly
affected by stellar evolution than primordial binaries, and a larger
fraction enter a phase of mass transfer.  The merger rates (per
binary) between main-sequence stars and (sub)giants, and between two
main-sequence stars, are considerably smaller for tidal binaries than
for primordial binaries (see Table~\ref{Tab_merge}), mainly because
these stellar types are the dominant constituents of primordial binary
systems.  For all other types of binary the tidal-binary merger rate
exceeds that for primordial binaries.  The merger rate between two
white dwarfs is enhanced due to the longer lifetime of the close
binary, during which time gravitational radiation can reduce the
orbital separation to the point of coalescence.

\begin{figure}
\hspace*{0.5cm}
\epsfxsize = 4.5cm
%\epsffile{fig_hrdCT.ps}
\psfig{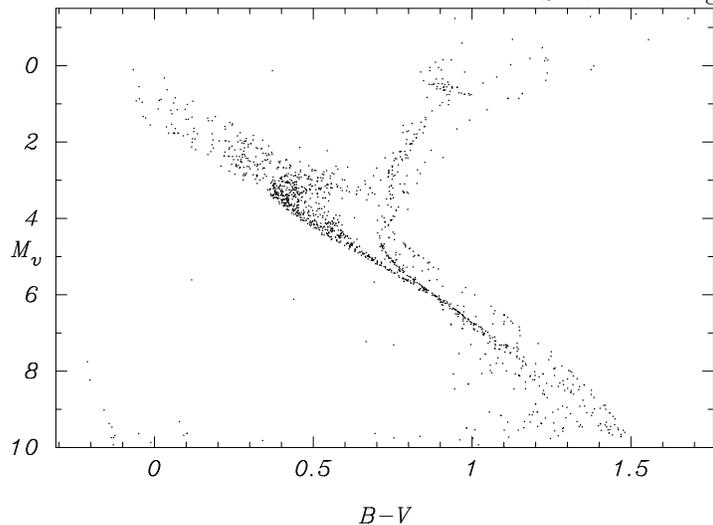}
\caption[]{Hertzsprung-Russell diagram of a population of tidal
binaries and single merger remnants, at a system age of
(model $C$) of 12~Gyr.}
\label{fig_hrdCT}\end{figure}

Figure~\ref{fig_hrdCT} shows the Hertzsprung-Russell diagram for the
tidal binaries of model $C$ at a system age of 12 Gyr.
The large diversity of objects is striking: blue stragglers as well as
many yellow stragglers and white dwarfs are visible. 

Figure~\ref{fig_nbss} shows the numbers of blue and yellow stragglers as
fractions of the total number of tidal binaries.  After a
few hundred million years of dynamical encounters, the blue and yellow
straggler populations become important components of the stellar
system.  After about 2~Gyr of dynamical encounters, the numbers of
blue and yellow stragglers reach equilibrium.  The fraction of blue
stragglers per tidal binary is roughly constant, at about 13\%, over
the full evolution time of the simulation, much larger than for single
stellar collisions or a population of primordial binaries.  The
start-up time of about 1~Gyr can be understood from the knowledge that
the mean tidal binary lifetime is about a Gyr (see
Table~\ref{Tab_blife} and Fig.~\ref{fig_binsurv}).

\begin{figure*}
\vspace*{-6.cm}
\hspace*{5.75cm}
\epsfxsize = 1.9cm
%\epsffile{fig_fateCT.ps}
%\hspace*{1.cm}
\psfig{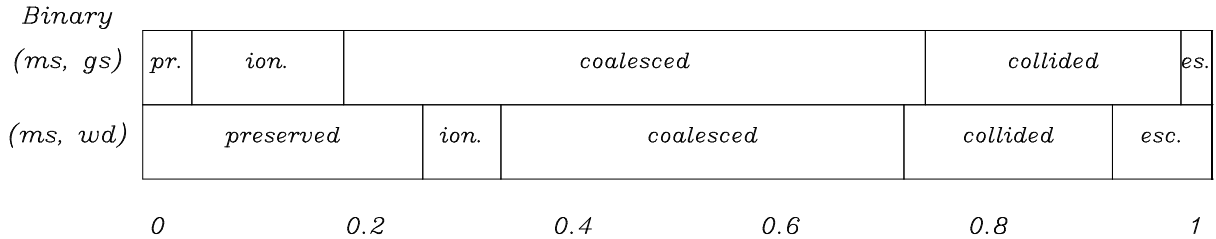}
\caption[]{Cannels (see text) by which a tidal
binary is destroyed (model C), for two types of tidal binaries.  The
upper bar gives the relative frequencies of tidal binaries where a
main-sequence star was captured by a (sub)giant or a main-sequence
star.  The lower bar indicates the relative frequency
of captured white dwarfs.  For details, see Table~\ref{Tab_term}.
}
\label{fig_fateCT}\end{figure*}

Figure~\ref{fig_fateCT} illustrates the various destruction channels
for a binary formed by the tidal capture of a main-sequence star by an
evolved star (ms, gs) or a white dwarf (ms, wd).  At the moment of
destruction, a large fraction of all tidal binaries in our simulations
still reflect their initial conditions.  (This is not true for
primordial binaries, whose appearance is changed dramatically by
exchange or collision-binary interactions; see
Tables~\ref{Tab_scatter} and \ref{Tab_term}).  Collision and
coalescence are common fates of tidal binaries, and the majority of
the (ms, gs) binaries end their lives as single objects.  Because of
the smaller physical sizes of the stellar components, (ms, wd)
binaries result are somewhat less likely to merge, and a larger
fraction of these binaries escape or survive to the end of the
calculation.

Once a white dwarf passes the Chandrasekhar limit, it is completely
destroyed in a type {\rm I}a supernova, which dissociates the binary.
The total supernova rate for tidal binaries is almost 0.08
supernovae per binary.  Primordial binaries contribute for about 0.01
(supernovae per primordial binary) to the total type {\rm I}a
supernova rate.  In the core of a high-density stellar system, the
total number of binaries formed by tidal capture can be estimated
using the data given in Table~\ref{Tab_init}.  With these numbers, we
arrive at a rate of about one type {\rm I}a supernova every 
9.28~Myr$/0.08 \simeq 120$~Myr.
This is about 30\% smaller than the supernova rate derived
for single star encounters in paper~I. The small contribution to the
type {\rm I}a supernova rate from tidal binaries results from the
capture frequency being smaller than the collision frequency.

\section{Results for the Salpeter mass function}
\label{sec_74}

In the model computation with a Salpeter mass function (model $S$ for
Salpeter, see paper~I and also Tables~\ref{Tab_initb} and
\ref{Tab_init}) we consider three different populations of binaries:
50000 non-dynamically and 5000 dynamically evolving primordial
binaries, all initialized at $t=0$ (sect.~\ref{sec_modelS_nd},
\ref{sec_modelS_d}), and tidal binaries
(sect.~\ref{sec_modelS_td}).  The masses of the primaries are chosen
between 0.1~\msun and 100~\msun, the other initial conditions are
summarized in Table~\ref{Tab_initb}.  Tidal binaries are
initialized as described in sect.~\ref{sec_pbin}, but due to the
small encounter rate in model $S$ (see paper~I) only 1000 tidal
binaries are modeled.

\subsection{Primordial binaries without encounters}
\label{sec_modelS_nd}

The steep mass function leads to many primaries having masses well
below the turn-off, even at a system age of 16~Gyr.  As a result, a
large fraction of the non-dynamically evolving binaries of model $S$
do not reach Roche-lobe contact during the simulation.  These binaries
are listed as (ms, ms) binaries in Table~\ref{Tab_blife}.  Binaries
containing a Wolf-Rayet star are rare, both due to the small number of
high-mass primaries, and to the short lifetimes of Wolf-Rayet stars.
The total number of black holes and neutron stars is negligible; this
is reflected in the low probability of finding one in a
non-dynamically evolving binary (Table~\ref{Tab_blife}).  
A binary in which a neutron star is formed is likely to be dissociated by the
asymmetry of the supernova, so few binaries contain neutron stars.
(The distribution of kick velocities in our model is described with
model $AK$ in Portegies~Zwart \& Verbunt 1996.)

The merger rate between white dwarfs is strongly enhanced in model $S$
relative to model $C$ (see Table~\ref{Tab_merge}).  This is mainly a
result of small differences in the initial mass-ratio
distribution. Although the initial distributions are chosen identical
(see Table~\ref{Tab_init}) the adopted rejection technique tends to
select equal mass pairs more often for model $S$ than for model $C$
(see sect.~\ref{sec_selecb}).  These binaries are more likely to form
(wd, wd) pairs which will subsequently emit gravitational waves,
ultimately resulting in the merger between the two stars.  The total
number of white dwarfs (single as well as in binaries) in model $C$ is
obviously higher than in model $S$ (see Table~\ref{Tab_blife}).

\subsection{Primordial binaries with encounters}
\label{sec_modelS_d}

The encounter rate between single stars and primordial binaries is
small, and due to the steep mass function, the majority of encounters
involve low-mass main-sequence stars.  The fractions of exchanges,
collisions and resonance encounters are consequently small compared to
those in model $C$.  The majority of binaries survive to the end of
the computation, and the average lifetime of a binary is almost as
large as in the non-dynamically evolving case (see
Table~\ref{Tab_blife}).  Ionization is the second most important
termination channel (see Table~\ref{Tab_term}).  The low encounter
rate is reflected in the small difference between dynamically and
non-dynamically evolving binaries (see Tables~\ref{Tab_blife} and
\ref{Tab_merge}).

\subsection{Tidal binaries}
\label{sec_modelS_td}

Because the low cluster density and small semi-major axes of tidally
formed binaries results in very low encounter rates in model S, we
have simply omitted these rates from Table~\ref{Tab_init}.  The
majority of tidal binaries survive throughout the evolution of the
stellar system without ever experiencing an encounter.
\section{Outlook}
\label{sec_75}

This is the second paper in our star cluster ecology series.  Our main
aim is to provide stepping stones toward a full integration of stellar
dynamics and stellar evolution for star cluster simulations.  Paper I
provided the first stepping stone: a treatment of collisions between
single stars, drawn from a prescribed initial mass distribution and
evolving independently between encounters.  In the present paper we
have taken the next step, incorporating a treatment of two more
complicated processes: 1) the formation of binaries by tidal capture;
2) collisions between single stars and primordial binaries.

Subsequent papers in this series will apply the techniques that we
have developed so far to full $N$-body systems.  In these more
realistic star cluster simulations, each star or binary, while
unperturbed, will evolve according to the prescriptions given by
Portegies Zwart \& Verbunt (1996).  The tools developed in this and
the previous paper will enable us to model stellar evolution during
periods of strong interactions.

In contrast to the following papers, the present paper is not meant to
provide results that can be directly compared to observations.  The
approximations made are too extreme, and the initial conditions too
simple, for such a comparison to be meaningful.  The main aim of the
present exercise is to give the reader confidence in and understanding
of the techniques we have developed.  The more realistic results from
future papers in this series can then be checked and extended
independently by others.  Rather than providing black boxes for the
dynamics and evolution of the stars in a cluster, our goal is to make
the modeling procedure transparent.

\acknowledgements
This work was supported in part by the Netherlands
Organization for Scientific Research (NWO) under grant PGS 78-277, by
the National Science Foundation under grants ASC-9612029 and
AST-9308005, and by the Leids Kerkhoven Boscha Fonds.  SPZ thanks the
Institute for Advanced Study and the University of Tokyo for their
hospitality.  Edward P.J.\ van den Heuvel of the Astronomical
Institute ``Anton Pannekoek'' is acknowledged for financial support and
for inviting our group for an extended work visit.

\bibliographystyle{aabib}
\bibliography{sdyn_II}

\end{document}